\documentclass[12pt]{scrartcl}
\usepackage[a4paper, margin=2.54cm]{geometry}
\usepackage[authoryear]{natbib}
\usepackage{amsmath, amssymb, amsthm}
\usepackage{graphicx}
\usepackage{tikz}
\usepackage{xcolor}
\usepackage{soul}
\usepackage{bbm}
\usepackage{enumerate}
\usepackage[shortlabels]{enumitem}
\usepackage{booktabs} 
\usepackage{caption}
\usepackage{subcaption}
\usepackage{float}
\usepackage{wrapfig}
\usepackage{csquotes}
\usepackage{changepage}
\usepackage{mathtools}
\usepackage{hyperref}

\newcommand{\calwidth}{\rho}
\newcommand{\ELPD}{\operatorname{ELPD}}
\newcommand{\scalingFactor}{\tau}

\title{Local prediction pools}
\author{Oscar Oelrich$^{a}$\thanks{Corresponding author: oscar.oelrich@stat.su.se}, Mattias Villani$^{a,b}$ and Sebastian Ankargren}
\date{%
    $^a$Department of Statistics, Stockholm University\\%
    $^b$Dept. of Computer and Info. Science, Link{\"o}ping University\\
    [2ex]%
}

\usepackage{fancyhdr}
\begin{document}

\maketitle

\begin{abstract}
    We propose local prediction pools as a method for combining the predictive distributions of a set of experts conditional on a set of variables believed to be related to the predictive accuracy of the experts. This is done in a two step process where we first estimate the conditional predictive accuracy of each expert given a vector of covariates---or pooling variables---and then combine the predictive distributions of the experts conditional on this local predictive accuracy. To estimate the local predictive accuracy of each expert, we introduce the simple, fast, and interpretable \emph{caliper method}. Expert pooling weights from the local prediction pool approaches the equal weight solution whenever there is little data on local predictive performance, making the pools robust and adaptive. We also propose a local version of the widely used optimal prediction pools. Local prediction pools are shown to outperform the widely used optimal linear pools in a macroeconomic forecasting evaluation, and in predicting daily bike usage for a bike rental company.\\
    \emph{Keywords:} Bayesian predictive synthesis; density forecasts; combining forecasts; macroeconomic forecasting; nonparametric methods.
\end{abstract}

\section{Introduction}

Forecast combination has a long history in statistics and related areas \citep{clemen_combining_1989,winkler_combining_1981} and is widely used in forecasting and policy making \citep{adolfson_modern_2007}. Early approaches focus on aggregating point forecasts \citep{bates_combination_1969}, whereas a more recent strand of the literature is more concerned with combining forecast distributions \citep{hall_combining_2007, geweke_optimal_2011, billio_time-varying_2013, kapetanios_generalised_2015, yao_using_2018, mcalinn_multivariate_2020, mcalinn_mixed-frequency_2021, casarin_flexible_2023}. These combined predictive distributions may come from statistical models learned from data, or be elicited directly from experts without explicit probabilistic models, or be a mix of the two types. 

An example of forecast combination is macroeconomic forecasting and policy making at central banks where predictive distributions from dynamic stochastic general equilibrium (DSGE) and vector autoregressive (VAR) models are combined with forecast distributions elicited from internal experts \citep{kjellberg_riksbanks_2010}. We will use the terms \emph{expert} and \emph{expert distribution} irrespective of whether the predictive distribution comes from statistical models or from elicitation of expert opinions.

The optimal way to linearly combine statistical predictive models is to jointly estimate the model parameters in all models and the pooling weights in the combined prediction in a mixture model \citep{fruhwirth-schnatter_finite_2006}. This ideal is often unattainable in practice however, either because the set of predictive distributions includes informally elicited expert opinion or because the models are too complex to be estimated jointly as a mixture. An example of the former is when large forecasting institutions use dedicated teams that work on models in isolation, using their own software implementations, which makes it practically impossible to re-estimate as a single mixture model. 
\citet{del_negro_dynamic_2014} call this informational friction.

A common approach to the combination of expert distributions in the literature is the \emph{linear prediction pool} \citep{lindley_reconciliation_1979, hall_combining_2007, geweke_optimal_2011} where the combined distribution is a linear, often convex, combination of the expert distributions. Such linear pools have been shown to be optimal from a Bayesian perspective under certain specific assumptions \citep{genest_combining_1986, west_modelling_1992}. The expert weights in linear pools are usually chosen to maximize the out-of-sample predictive performance with respect to some scoring rule, most often the logarithmic scoring rule; such optimized pools have been termed \emph{optimal prediction pools}. A related set of aggregation methods are called stacking in the machine learning literature \citep{wolpert_stacked_1992} and have more recently also been further developed in the statistical literature \citep{yao_using_2018}. \cite{geweke_optimal_2011} show that optimal prediction pools will typically converge to a solution that puts non-zero weight on more than one model in large samples; this is in contrast to Bayesian model averaging where the posterior model probabilities will asymptotically concentrate entirely on one of the models, even when the data generating process is outside the set of compared models \citep{berk_limiting_1966}. 

The original linear and optimal prediction pools use a single time-invariant weight for each model. We will term such a weighting scheme a \emph{global pool}. Global pools implicitly make the strong assumption that the predictive ability of the experts is the same over time and for all possible values of any conditioning variables used as explanatory variables in the models. Some recent work have proposed to use time-varying weights in optimal prediction pools to allow models to be up- or down-weighted during certain time periods, see e.g. \citet*{del_negro_dynamic_2014} and \citet*{billio_time-varying_2013}. \citet{li_bayesian_2022} have recently proposed a generalization of the optimal prediction pools in \citet{geweke_optimal_2011} where the weights are allowed to depend on a set of covariates through a softmax function. Similarly, \citet{yao_bayesian_2021} extends the stacking method of \citet{yao_using_2018} by allowing the model weights to vary as a function of the data.

In this paper, we take a general perspective similar to that in \citet{yao_bayesian_2021} and allow the expert weights to vary with respect to a general set of \emph{pooling variables}, which are variables that are believed to affect the predictive ability of the experts. These pooling variables may include time---giving us time-varying expert weights---but also other variables that may be related to expert performance. The pooling variables may be part of the information set of some of the experts, but can equally well be completely external variables not used by any of the experts, for example a business cycle indicator aggregated from survey expectations or sentiments extracted from social media. We call such weighting schemes \emph{local prediction pools} to emphasize that they are determined by the local predictive performance of the experts.

The main challenge with local prediction pools is the need to learn the local predictive performance of all the experts. The learned local performance must allow for robust interpolation and extrapolation across the space of the pooling variables for it to be useful when constructing pools for predicting new data. This is a challenging problem, particularly when the number of pooling variables is large and limited data is available on the prediction performance of the experts.

Given historical measures of predictive ability for each expert and data on the pooling variables, learning local prediction performance is a problem of surface estimation. The \emph{pooling surface} for each of the experts can be estimated using a multitude of smoothing techniques where the pooling surface is estimated by averaging locally around the point of interest in the space of pooling variables. With this perspective, a global prediction pool is an extreme special case where all observations are used equally to estimate global performance and to construct a single weight on each model,  regardless of the state of the local pooling variables. We propose an easily implemented nonparametric method for estimating the pooling surface that automatically adapts the degree of locality to the local concentration of data in the pooling space and the differing historical local performance of the models in the pool. The expert weights from this estimator approaches equal weights locally as the number of past local predictions decreases. We also introduce a local version of the optimal prediction pool in \citet{geweke_optimal_2011}.

To allow us to interpret local prediction pools in subjectivist Bayesian terms we take the decision maker perspective of \citet{lindley_reconciliation_1979}, where expert predictions are treated as data used by a \emph{decision maker} to update her predictive beliefs. We formalize our local prediction pools using an extension of the Bayesian synthesis framework in \citet{johnson_bayesian_2018}.

The rest of the paper proceeds as follows: Section \ref{framework} develops the local prediction pools framework; Section \ref{caliper} introduces the caliper method for estimating local predictive ability together with an illustrative theoretical example; Section \ref{applications} contains two applications, in the first we use local prediction pools to make better quarterly forecasts of key macroeconomic variables, and in the second we predict daily bike usage for a bike rental service; Section \ref{conclusions} concludes.

\section{The local pooling framework}\label{framework}

This section establishes a theoretical framework for local prediction pools in which a \emph{decision maker} ($\mathcal{DM}$) wants to create a combined, or \emph{pooled}, predictive distribution for a variable of interest, $y_{t}$, based on the predictive distributions of $K$ \emph{experts}. The experts may be formal statistical models or opinionated humans. To help accomplish this, the decision maker uses historical data in the form of a sequence of predictions made by the experts. Furthermore, the $\mathcal{DM}$ also has access to a vector of \emph{pooling variables}, $\mathbf{z}_t \in \mathcal{Z}$, over which she believes that the predictive ability of the experts vary. The aim of the decision maker is  to pool the experts' forecasts based on their local predictive ability at the current $\mathbf{z}_t$. 

To achieve her aim, the decision maker needs to i) set up a pooling space $\mathcal{Z}$, ii) estimate the local predictive ability of each expert over $\mathcal{Z}$, and iii) use a pooling function to synthesize the predictions of the experts, conditional on their local predictive ability. This section goes through these steps in turn, and positions local pools within the Bayesian predictive synthesis (BPS) framework of \citet{johnson_bayesian_2018}.

Any scoring function can be used to measure the predictive ability in step ii) but we will use the logarithmic scoring rule in the form of the log predictive density. The logarithmic scoring rule has the unique advantage of being both local and proper \citep{bernardo_bayesian_1994}, and is commonly used in model selection. Further, the \emph{expected log predictive density} (ELPD) of a predictive distribution is proportional to the Kullback-Leibler divergence with regards to the data-generating process \citep*{hall_combining_2007}. The linear pooling function in step iii) is motivated by the \emph{Bayesian predictive synthesis} framework \citep{johnson_bayesian_2018}.

\subsection{Setting up the pooling space}

The pooling space $\mathcal{Z}$ should include all variables that the $\mathcal{DM}$ believes co-vary with the experts' predictive ability. This can include (transformations of) covariates used by the experts, as well as variables that none of the experts use. While the $\mathcal{DM}$ will often include some, or even all, of the covariates used by the experts, this does not have to be the case. In theory, it is possible to set up a pooling space without even knowing which variables the experts used when they produced their forecasts. The space $\mathcal{Z}$ should be constructed using variables that the $\mathcal{DM}$ \emph{perceives} as determinants of local predictive ability.
 
\bigskip

\subsection{Estimating local predictive ability over $\mathcal{Z}$}

The purpose of using a local pool is to exploit variations in predictive ability over $\mathcal{Z}$. We conceptualize this variation as a hypersurface in $\mathcal{Z}$ for each expert. An intuitive measure of predictive ability of a model is the expected log predictive density (ELPD) for a new data point \citep{gelman_understanding_2014}. The ELPD of expert $k$, trained on a sample $(y_1, ..., y_T)$, for a new single observation from the data-generating process is given by
\begin{equation}
    \operatorname{ELPD}^{(k)} = \int 
    \log p_{{k}} \left( 
        \tilde y_{{T+1}} \mid y_{1}, \dots, y_{T}
    \right)
    dF(\tilde y_{{T+1}}),
\end{equation}
where $p_{{k}} \left( 
        \tilde y_{{  T}+1} \mid y_1, \dots, y_{  T}
    \right)$ is the predictive distribution of expert $k$ and $F(\tilde y_{{  T+1}})$ is the cdf of the data-generating process. We denote the \emph{local} expected log predictive density of a model $k$ for a specific point $\mathbf{z}_{T+1}$ in $\mathcal{Z}$ by
\begin{equation}
    \operatorname{ELPD}^{(k)}(\mathbf{z}_{  T+1}) = 
    \int \log p_{k}
    \left( 
        \tilde y_{{  T}+1} \mid y_1, \dots, y_{  T}
    \right)
    dF(\tilde y_{{  T}+1} \vert \mathbf{z}_{{  T}+1}).
\end{equation}

Estimating local predictive ability is a challenging problem since the predictions of the experts are typically sparse in $\mathcal{Z}$, especially when $\mathcal{Z}$ is high-dimensional. To tackle this estimation problem, the $\mathcal{DM}$ is free to use whatever parametric or non-parametric model she thinks best captures how the predictive abilities of the experts change over $\mathcal{Z}$. This can mean simple parametric regression models, more elaborate modeling of smoothness using Gaussian processes, or non-parametric techniques like $k$-nearest neighbors.

Local predictive ability does not have to be modeled in the same way for all experts. This allows the decision maker to incorporate beliefs about generalizability that differs between experts. For example, the predictive ability of a complex model might vary more quickly over $\mathcal{Z}$, and  the decision maker may therefore be less certain about the predictive performance for regions in $\mathcal{Z}$ that the model has not visited in the past.


\subsection{Synthesizing predictive distributions}

The final step in forming the local prediction pool is the combination of the predictions made by the experts, conditional on their (estimated) local predictive ability. Exactly how this combination is to be done is ultimately up to the $\mathcal{DM}$, but we will limit the scope of this paper by only considering linear pools of the form
\begin{equation}
   p_{{\scriptscriptstyle \mathcal{DM}}}(y_{t+1}\mid \mathcal{H}) 
    = w_1 p_1(y_{t+1}) + \dots + w_K p_K(y_{t+1}),
\end{equation}
where $\mathcal{H}$ denotes the set of historical predictive distributions supplied by the $K$ experts, $p_k$ is the predictive distribution of expert $k$ for $y_{t+1}$, and $w_k$ is the weight given to that same expert. A reasonable constraint to put on the weights is to have them be non-negative and summing to one, as in the optimal linear prediction pools of \citet{hall_combining_2007} and \citet{geweke_optimal_2011}, where the weights of the experts are selected to maximize the historical performance of the pool. Linear pools are simple yet powerful, and have the additional advantage of allowing us to reframe the third step in subjectivist Bayesian terms as Jeffrey's updating \citep{johnson_bayesian_2018}.

The problem of how to combine conflicting probability assessments, such as predictive distributions, has a long history \citep{lindley_reconciliation_1979}. One solution is the decision maker approach where the predictive distributions are treated as data to be used by a decision maker \citep{genest_combining_1986}. Once the distributions are taken as data points, it becomes fairly straight-forward to think in conventional Bayesian terms of prior to posterior updating. 

\citet{johnson_bayesian_2018} show that the use of linear pools can be justified from a subjective Bayesian perspective through a framework they call Bayesian predictive synthesis (BPS). BPS uses a synthesis function that specifies the posterior conditional on the predictions of the experts. 
\begin{equation}
    p(y\mid \mathcal{H}) = \int \alpha(y|\boldsymbol{x}) h(\boldsymbol{x})\,dh(\boldsymbol{x}),
\end{equation}
where $\mathcal{H}=\big(h_{1}(\cdot),\ldots,h_{K}(\cdot)\big)$ is the set of predictive densities supplied by the experts and $\alpha(y|\boldsymbol{x})$ is the synthesis function.
This updating does not obtain the posterior through the application of Bayes theorem but rather through Jeffrey's updating \citep{diaconis_updating_1982}. 

\citet{johnson_bayesian_2018} derive a linear pool version of BPS
\begin{align}
    p(y\mid \mathcal{H}) = 
        \int 
            \left(
                \sum_{k=1}^K w_k \delta_{x_k}(y) 
            \right)
            h(\boldsymbol{x})\,dh(\boldsymbol{x})
        =
        \sum_{k=1}^K w_kh_k(y),\label{johnson_west_linear}
\end{align}
where $\delta_x(y)$ is the Dirac delta function. We can easily extend \eqref{johnson_west_linear} to a local pool by letting the weights depend on a vector of pooling variables $\mathbf{z}$
\begin{equation}\label{BPS_extend}
    p(y\mid \mathcal{H}, \mathbf{z}) = 
        \int 
            \left(
                \sum_{k=1}^K w_k(\mathbf{z}) \delta_{x_k}(y) 
            \right)
            h(\boldsymbol{x})\,dh(\boldsymbol{x})
        =
        \sum_{k=1}^K w_k(\mathbf{z})h_k(y),
\end{equation}
where $\mathcal{H}$ is the set containing the $K$ predictive distributions supplied by the experts. This extension allows us to position local prediction pools within the Bayesian predictive synthesis framework. 

\section{The caliper method for learning local predictive performance}\label{caliper}

In this section we propose the \emph{caliper method}  as a simple, interpretable way of modeling local predictive ability and combining expert forecasts in a linear pool. We use a simulated example to illustrate the method.

\subsection{The caliper method}\label{subsec:calipermethod}
The caliper method estimates $\ELPD(\mathbf{z})$ by averaging all historical log predictive scores that occurred within a given distance (caliper width) from $\mathbf{z}$. Formally, the decision maker estimates the local $\ELPD(\mathbf{z})$ for expert $k$ by
\begin{equation}
    \widehat{\operatorname{ELPD}}^ {(k)}(\mathbf{z}) = \frac{1}{{n_{\calwidth}(\mathbf{z})}}\sum_{i \in \mathcal{I}_{\calwidth}(\mathbf{z})} \log p_{(k)}(y_i),
\end{equation}
where $\mathcal{I}_{\calwidth}(\mathbf{z})$ is the set of $n_{\calwidth}(\mathbf{z})$ observations that lie within a caliper of width $\calwidth$ centered at $\mathbf{z}$. We will use the Euclidean distance on standardized pooling variables in the applications, but any distance measure can be used to define the caliper. When $n_{\calwidth}(\mathbf{z})=0$, i.e. when there are no historical observations within the caliper, the $\ELPD(\mathbf{z})$ estimate is set to zero for each expert, leading to equal weights when combining predictions.

The caliper method is similar to $k$-nearest neighbors (kNN). However, there are two important differences:
\begin{enumerate}[a)]
    \item kNN will always base its estimate on the $k$ nearest observations, regardless of distance. If all observations are far away, the kNN estimate can therefore be based on data of dubious relevance. The caliper method, on the other hand, will only include observations it regards as close enough, and will default to equal weights when there is no relevant data.
    \item kNN will use exactly $k$ observations, even when there are many more observations close by. The caliper method, on the other hand, is capable of exploiting variation in the data density in $\mathcal{Z}$.
\end{enumerate}

Once the decision maker has access to estimates of local predictive ability for each expert, she needs to combine these predictions in some way. The caliper method combines predictive distribution using a local linear pool 
\begin{equation}
    p(y \mid \mathcal{H}, \mathbf{z}) = \sum_{k=1}^{K} w_k(\mathbf{z}) p_k(y)
\end{equation}
where the weight of expert $k$ is calculated by feeding the estimates of local predictive ability through a softmax transformation:
\begin{equation}\label{eq:naturalscaling}
    w_k(\mathbf{z}) = \frac{
            \exp\Big(n_\rho(\mathbf{z}) \times \widehat{\operatorname{ELPD}}^ {(k)}(\mathbf{z})\Big)
        }{
            \sum_{j=1}^{K} \exp\Big(n_\rho(\mathbf{z}) \times \widehat{\operatorname{ELPD}}^ {(j)}(\mathbf{z})\Big)
        },\quad k=1,\dots,K,
\end{equation}
The weights in \eqref{eq:naturalscaling} use what we will refer to as \emph{natural scaling} where the local ELPD estimates are scaled by the number of observations, $n_\rho(\mathbf{z})$, used in forming the estimate. Natural scaling will lead to model weights that discriminate more sharply between models locally when there is more data available; Bayesian model averaging has the same behavior, but globally.

The caliper width, $\calwidth$, determines how close in $\mathcal{Z}$ a previous prediction has to be in order to be deemed relevant for the local estimates, the caliper width should therefore match how quickly the $\mathcal{DM}$ thinks ELPD changes over $\mathcal{Z}$. Selecting the caliper width is a question of bias-variance tradeoff: a smaller width will better capture the \emph{local} part of $\ELPD(\mathbf{z})$, but this will come at the expense of basing the estimate on fewer observations, thereby increasing variance. How small a caliper width the $\mathcal{DM}$ can afford will depend on the sample size and the dimension of $\mathcal{Z}$.

Natural scaling introduces a tension between the \emph{locality} of experts' performance and the degree of discrimination between experts: increasing the caliper width $\calwidth$ does not only affect the bias-variance trade-off in the locality of the estimate, it also changes the degree of discrimination between models. This means that the caliper width that gives best predictive performance may have little to do with how quickly predictive ability varies in $\mathcal{Z}$. To break this tension, we allow for departures from natural scaling by introducing a separate scaling factor $\scalingFactor$ in the softmax weights
\begin{equation}
    w_k(\mathbf{z}) = \frac{
            \exp\Big(\scalingFactor \times \widehat{\operatorname{ELPD}}^ {(k)}(\mathbf{z})\Big)
        }{
            \sum_{j=1}^{K} \exp\Big(\scalingFactor \times \widehat{\operatorname{ELPD}}^ {(j)}(\mathbf{z})\Big)
        },\quad k=1,\dots,K,
\end{equation}
The scaling factor determines how sharply we discriminate between models with differing estimated predictive ability; it allows us to modify the behavior of the synthezising step from equal weights ($\scalingFactor = 0$) to turning the synthesis into model selection ($\scalingFactor \rightarrow \infty$).

To use the caliper method with discrimination, the decision maker must specify two hyperparameters: i) the caliper width $\calwidth$ and ii) the scaling factor $\scalingFactor$. The $\mathcal{DM}$ could in principle put a prior on $\rho$ and $\scalingFactor$, but for the sake of simplicity we treat them as fixed hyperparameters for the decision maker to select. Alternatively, if the decision maker has no strong preferences for these hyperparameters, they can be determined by optimization. This would mean generating pooled predictions for a grid of values of $(\calwidth, \scalingFactor)$ and at time $t$ selecting the hyperparameters that gave best predictive performance in time periods before time $t$.

\bigskip

\subsection{Illustrative example}\label{illustration}

To illustrate the process of using local predictive pools and the caliper method, we work through an example in which the decision maker has access to predictions from two experts, each in the form of a model

\begin{align}
    \text{Expert 1}: \quad y &= 
        \alpha^{(1)} + \beta^{(1)} x_1 + \epsilon^{(1)},
        \hspace{0.5cm}
        \epsilon^{(1)}
        \overset{\mathrm{iid}}{\sim}
        \mathrm{N}(0,\sigma^{(1)}) \\
    \text{Expert 2}: \quad y &= 
        \alpha^{(2)} + \beta^{(2)} x_2 + \epsilon^{(2)},
        \hspace{0.5cm}
        \epsilon^{(2)}
        \overset{\mathrm{iid}}{\sim}
        \mathrm{N}(0,\sigma^{(2)}).
\end{align}
Each expert uses a diffuse normal-inverse-gamma (NIG) prior for the parameters to produce a Bayesian predictive distribution. In order to be able to generate example data, as well as to derive theoretical quantities like local predictive ability, we need to assume a specific data-generating process. We use the simple linear model
\begin{equation}\label{DGP}
    \mathrm{DGP}: \quad y = x_1 + x_2 + \epsilon,
\end{equation}
where $\epsilon \sim \operatorname{N}(0, 1)$ and new observations from the DGP are generated by independently drawing values of $x_1$ and $x_2$ from the $\operatorname{N}(0, 1)$ distribution.

The first step in creating a local prediction pool is for the $\mathcal{DM}$ to set up the pooling space. She decides to include the covariates of both experts in  $\mathcal{Z}$, so that $\mathbf{z}_T = (x_{1,T}, \, x_{2,T})$. If she wanted to expand $\mathcal{Z}$ she could include, for example, an interaction effect $(x_3 = x_1 \times x_2)$, higher order terms $(x_4 = x_1^2, \, x_5 = x_2^2)$, or a variable that neither expert uses.

\begin{figure}[ht]
        \centering
        \minipage{0.32\textwidth}
        \includegraphics[width=\linewidth]{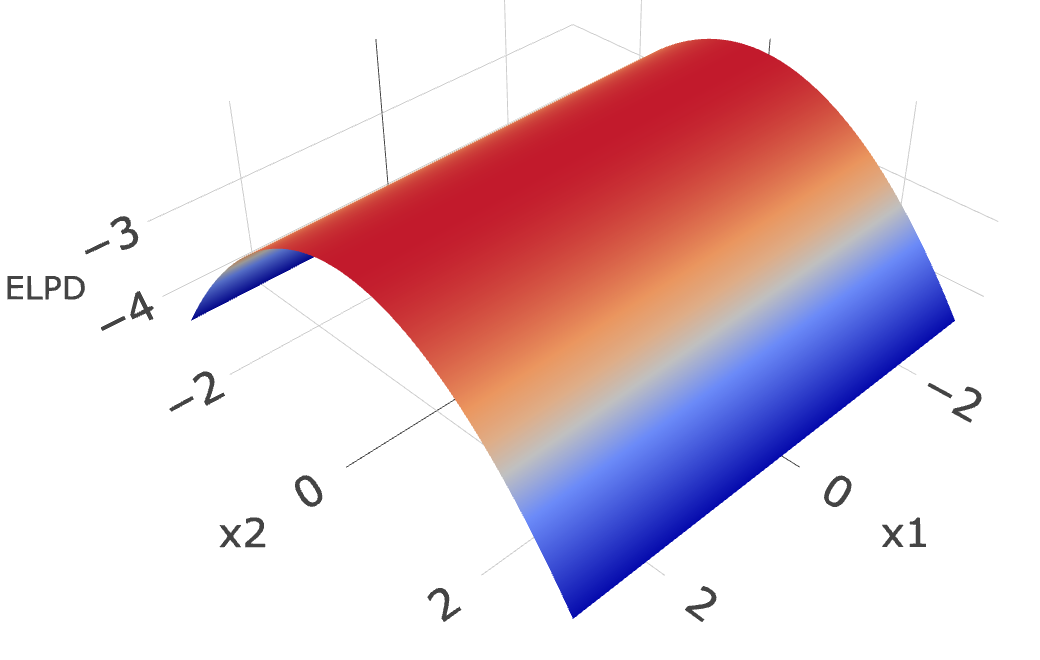}
        \caption*{a) Expert 1 ($x_2$ omitted)}\label{fig:lelpd1}
        \endminipage\hfill
        \minipage{0.32\textwidth}
        \includegraphics[width=\linewidth]{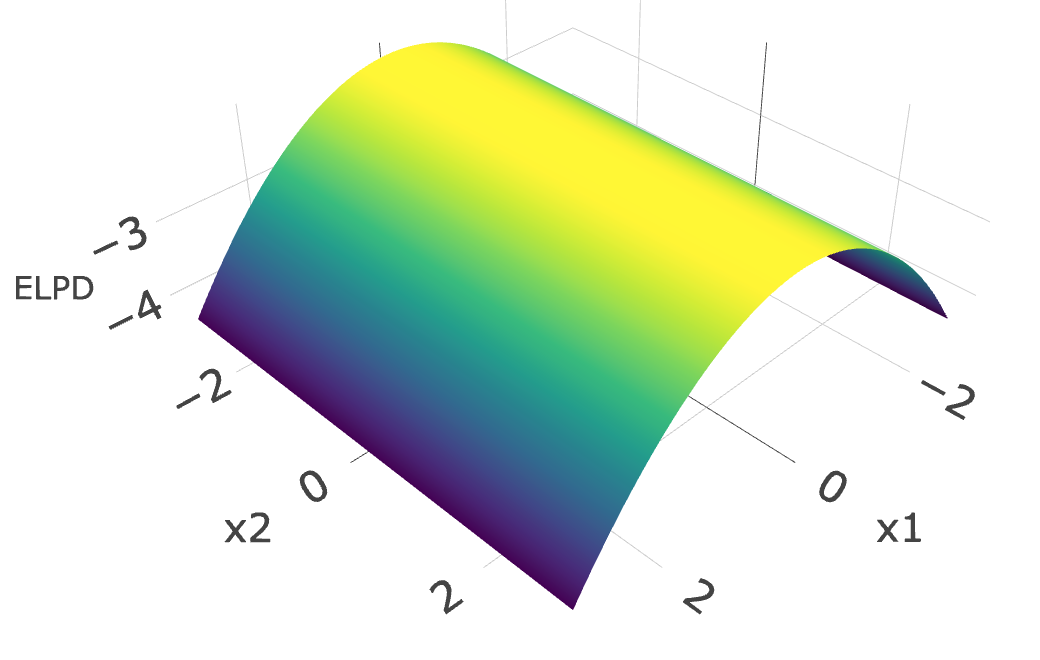}
        \caption*{b) Expert 2 ($x_1$ omitted)}\label{fig:lelepd2}
        \endminipage\hfill
        \minipage{0.32\textwidth}%
        \includegraphics[width=\linewidth]{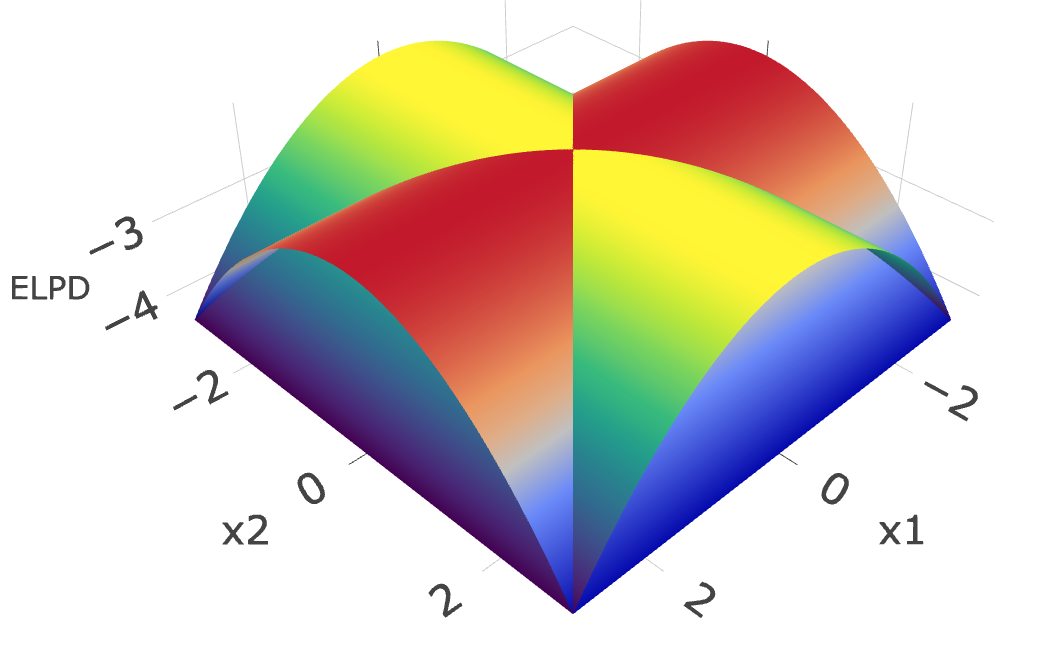}
        \caption*{c) Both experts}
        \endminipage
        \caption{Hypersurfaces of the predictive ability of the two experts in $\mathcal{Z}$.}\label{fig:lelpd12}
    \end{figure}

The second step in creating a local prediction pool is estimating the local predictive ability of each expert. As we have access to the data-generating process we can visualize how the predictive ability of the experts varies over $\mathcal{Z}$. The $\ELPD(\mathbf{z})$-surfaces of the experts can be found in Figure \ref{fig:lelpd12} a)--b). Each expert (unknowingly) omits one of the covariates in the DGP, and so the predictive ability of each expert deteriorates with the absolute value of this omitted covariate. Figure \ref{fig:lelpd12} c) illustrates that there are regions of $\mathcal{Z}$ where the predictive ability of one expert dominates. Using local pooling, we aim to capture this variation in predictive ability as a function of the pooling variables.

\bigskip
    
In most applications, the decision maker will not have access to the data-generating process, making it impossible to directly calculate how the predictive ability of each expert varies over $\mathcal{Z}$, and it therefore has to be estimated. In this example, the decision maker will use the caliper method with natural scaling, described in the previous section.

\begin{figure}[H]
    \centering
    \includegraphics{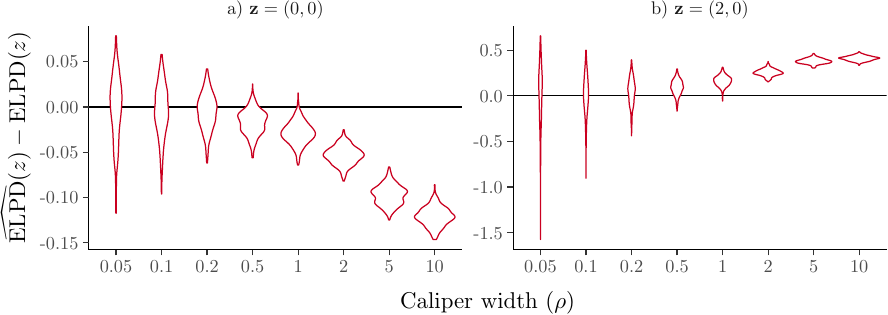}
    \caption{Sampling distribution of the errors in the estimate of local predictive ability of Expert 2 by the caliper method at two points in $\mathcal{Z}$.}
    \label{localELPDest_sim}
\end{figure}

The caliper method requires selecting a caliper width to control the inherent bias-variance tradeoff in estimating local predictive performance. Figure \ref{localELPDest_sim} shows the sampling distribution of the error in the estimate of $\ELPD(\mathbf{z})$ for Expert $2$ as a function of the caliper width. The figure was constructed by repeatedly sampling realizations of size $N=2000$ from the data generating process in \eqref{DGP} with the last $1000$ observations being used to estimate the predictive ability of the model. Since the predictive distribution of the expert depends on the realized data, each realization has its own true $\ELPD(\mathbf{z})$. 

Figure \ref{localELPDest_sim} a) illustrates the performance of Expert $2$ at the point $\mathbf{\mathbf{z}} = (0,0)$ where this expert fits the data well. Increasing the caliper width will lead to reduced variance, but also an increasing negative bias in the $\ELPD(\mathbf{z})$ estimate. This is because as we move further away from the point $(0,0)$ the caliper will cover areas where the model has worse fit than at $(0,0)$.

Figure \ref{localELPDest_sim} b) shows the performance of Expert $2$ at the point $\mathbf{\mathbf{z}} = (x_1, x_2) = (2,0)$. At this point Expert $2$ fits the data poorly since it omits $x_1$. Increasing the caliper width again leads to reduced variance, but now the bias in the $\ELPD(\mathbf{z})$ estimate will be increasingly positive. This is because the majority of new observations captured by the increasing caliper width will be from areas in $\mathcal{Z}$ where the model has better fit than at the current point.

The third step in creating a local prediction pool is aggregating the predictive distributions of the experts based on their (estimated) local predictive ability. In our example, the $\mathcal{DM}$ wants to make predictions at the two points $\mathbf{z} = (0,0)$ and $\mathbf{z} = (2,0)$. We compare the performance of the $\mathcal{DM}$ with two reference methods: a pool with equal weights, and the linear prediction pool of \citet{geweke_optimal_2011}.

\begin{figure}[H]
    \includegraphics{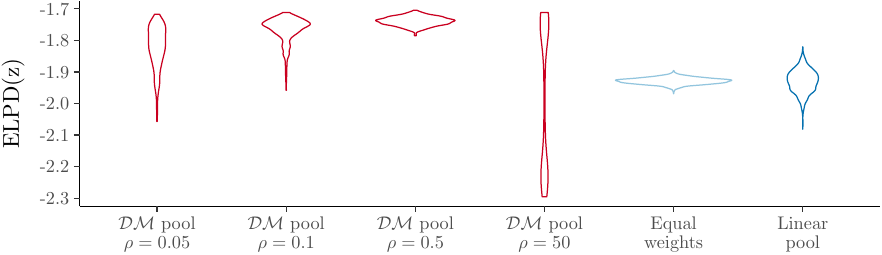}
    \caption{Expected log predictive density for a new observations at $\mathbf{z}=(2, 0)$ for the caliper method together with reference methods. Expectation is taken with regards to the data-generating process in \eqref{DGP}. (See the main text for details.)}
    \label{Default_lab}
\end{figure}

At $\mathbf{z} = (0,0)$ both models are equally misspecified, and make almost identical predictions. As each expert makes more or less identical predictions, any linear combination of their predictions will also be more or less identical. At $\mathbf{z} = (2,0)$ Expert 1, which includes $x_1$, greatly outperforms Expert 2. The caliper method captures this, which translates into markedly better predictions for a range of caliper widths. 

If we keep increasing the caliper width we will eventually arrive at an estimate of local predictive ability that is no longer local in any meaningful sense. For example, using the data-generating process in this simulation, a caliper width of $\calwidth = 50$ will almost always includes all previous observations. When this is the case and both of the experts have the same global predictive ability, we observe the same polarizing behavior as that of Bayesian posterior probabilities described in \citet{yang_bayesian_2018}. Since both models are equally misspecified globally over $\mathcal{Z}$, the difference in estimated predictive ability follows a random walk and will not converge to zero. As the sample size increases for any given sample, one of the models will therefore completely dominate the pool. Note that this is only the case for natural scaling.

\section{Applications}\label{applications}

In the applications we will refer to the local pooling method described in Section \ref{framework} as a local decision maker pool (\emph{local $\mathcal{DM}$} for short). We will also consider a pool that assigns equal weights to all predictive distributions (\emph{equal weights}) and the linear pool of \citet{geweke_optimal_2011}, which we will refer to as a global optimization-based pool (\emph{global opt.}), since it obtains its weights by optimizing the historical log scores over all of $\mathcal{Z}$.

The caliper method works by subsetting the data set based on variables that the decision maker believes that the predictive ability of the experts may vary over. This suggests that we could extend the global optimization-based linear pool into a local pool in a similar manner. To this end we introduce the local optimization-based linear pool (\emph{local opt.}), which works exactly as the pool in \citet{geweke_optimal_2011}, except that when optimizing the weights at time $t$, it only includes past predictions made within a given caliper width of $z_t$. If there are no past predictions over which to optimize, each expert is given the same weight.

\subsection{US macroeconomic forecasting}\label{empirical_macro}

In our first applied example we use the framework developed in the previous sections to forecast key macroeconomic time series in the US. The dataset used by the experts consists of the seven US macroeconomic variables in \citet*{smets_shocks_2007}: quarterly real GDP growth (\texttt{gdp}), quarterly inflation rate (\texttt{tcpi}), the federal funds rate (\texttt{fed}), quarterly real consumption growth, quarterly real investment growth, hours worked, and real compensation per hour. These time series are transformed in accordance with \citet{gustafsson_bayesian_2023}. 

The decision maker is interested in predicting the three variables \texttt{gdp}, \texttt{fed}, and \texttt{tcpi}. To aid her in this, the decision maker has access to experts in the form of predictive distributions from a set of models: i) a Bayesian homoscedastic VAR(1) model estimated on all seven variables, ii) a Bayesian VAR(1) model with stochastic volatility estimated on all variables, iii) Bayesian Additive Regression Tree models (BART, \cite{chipman_bart_2010}) for each of \texttt{gdp}, \texttt{fed}, and \texttt{tcpi} as univariate response variables with one lag of all seven macro variables as explanatory variables, and iv) a Bayesian VAR(1) model with time-varying parameters and stochastic volatility for the three-dimensional response vector with \texttt{gdp}, \texttt{tcpi} and \texttt{fed}. For each model class, we obtain the univariate one-step-ahead predictive distribution for \texttt{gdp}, \texttt{fed}, and \texttt{tcpi}; this allows us to explore differences in the local weighting schemes across the three variables. 

When setting up her pooling space $\mathcal{Z}$, the decision maker has access to all the variables used by the experts. In addition, the decision maker has access to an additional pooling variable in the form of ISM's manufacturing purchasing managers' index (\texttt{pmi}), which is not used by any of the experts \citep{lahiri_nowcasting_2013}. The data set includes $218$ observations, $72$ of which are used in the initial estimation of the experts' models. Using all eight variables to form $\mathcal{Z}$ is therefore not a good idea, as the $\mathcal{DM}$ would be estimating a hypersurface in an eight-dimensional space based on roughly $150$ observations. The decision maker therefore only uses GDP growth, inflation, the federal funds rate, and \texttt{pmi} to construct $\mathcal{Z}$.

\begin{figure}
    \centering
    \minipage{0.5\textwidth}
    \includegraphics{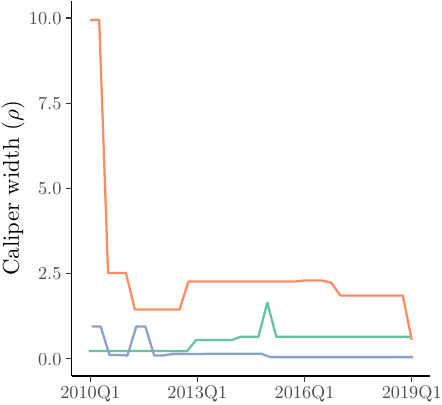}
    \caption*{a) Local $\mathcal{DM}$ pool.}
    \endminipage
    \minipage{0.5\textwidth}
    \includegraphics{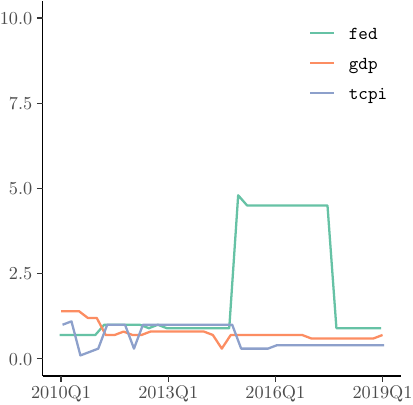}
    \caption*{b) Local opt. pool.}
    \endminipage
    \caption{Dynamically selected caliper widths for the macroeconomic data.}
    \label{cal_dyn_macro}
\end{figure}

The decision maker uses the caliper method with natural scaling to estimate local predictive ability in $\mathcal{Z}$---for a version using the caliper method with discrimination, see Appendix A. She continually updates the caliper width, $\calwidth$, at each time step by maximizing the historical log predictive density score over all previous aggregate predictions. The same method is used to dynamically select the caliper width of the local optimization-based pool. The dynamically selected caliper widths are shown in Figure \ref{cal_dyn_macro}.

\begin{figure}
    \centering
    \includegraphics{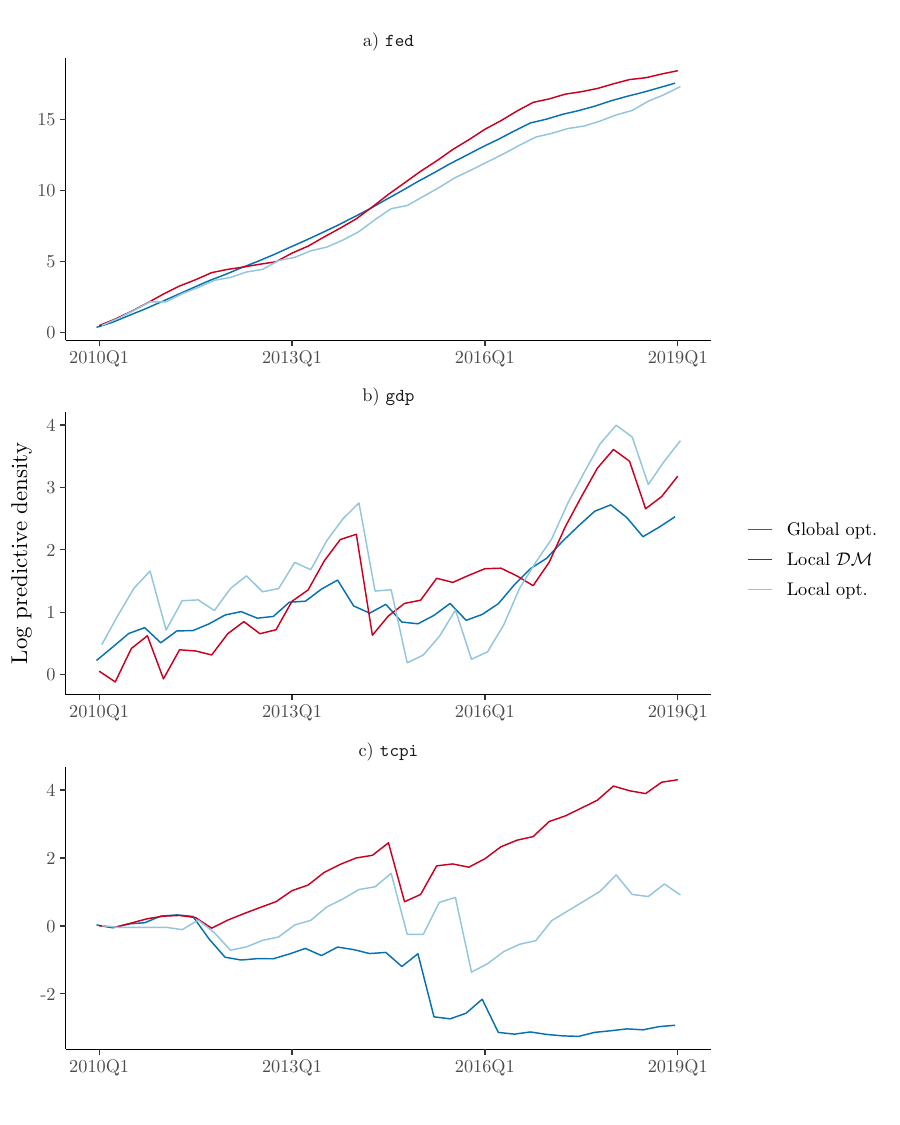}
    \caption{Cumulative log scores relative to equal weights of one-step-ahead quarterly forecasts.}
    \label{macro_cumu}
\end{figure}

Table \ref{table:all} displays the sum of out-of-sample log predictive scores for all the methods, and Figure \ref{macro_cumu} shows the development of these log scores over time relative to the equal weights method. All methods outperform equal weights by roughly the same amount when predicting \texttt{gdp} and \texttt{fed}. For \texttt{tcpi}, the local $\mathcal{DM}$ pool performs the best, outperforming the local optimization-based pool with some margin. All aggregation methods outperform the best individual experts. See \citet{villani_regression_2009} for a discussion of how differences in the log predictive scores can be loosely interpreted using Jeffreys' scale of evidence for log Bayes factors.

\begin{table}[b]
    \centering
    \begin{tabular}{lrrrr}
    \toprule
    & Equal weights & Global opt. & Local opt. & Local $\mathcal{DM}$ \\
    \midrule
    \texttt{fed}  & 34.4  & 51.9  & 51.7 & \textbf{52.8}  \\ 
    \texttt{gdp}     & -31.4 & -28.9 & \textbf{-27.6} & -28.2       \\ 
    \texttt{tcpi}  & -20.4 & -23.3 & -19.5 & \textbf{-16.1}\\ 
    \bottomrule
    \end{tabular}
    \caption{Comparison of different pooling schemes. Sum of log predictive densities for one-step-ahead quarterly forecasts of the three variables \texttt{fed}, \texttt{gdp}, and \texttt{tcpi}, for the period 2010:Q1 to 2019:Q1. Bold numbers indicate the best method for each variable.}
    \label{table:all}
\end{table}

\begin{figure}
    \centering
    \includegraphics{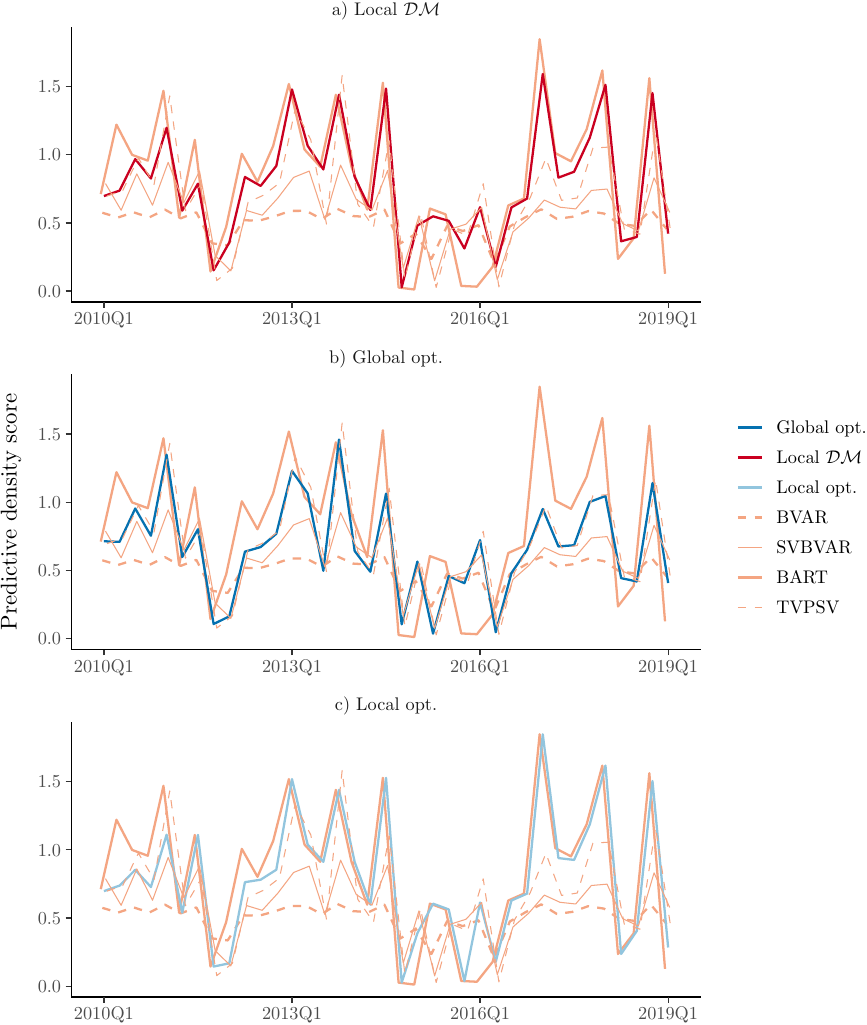}
    \caption{Predictive density scores of one-step-ahead, quarterly, predictions of \texttt{tcpi}. Each subplot relates one of three pooling methods to the four experts.}
    \label{tcpi}
\end{figure}

Figure \ref{tcpi} explains why the two local pools outperform the globally optimized pool for \texttt{tcpi} by displaying the log predictive density evaluations over time for the four pooling schemes and the individual experts. The figure shows that while the globally optimized pool relies almost exclusively on the TVPSV model, which has the best performance over the whole data set, the local pools correctly put greater weight on the BART model when it performs well, and opts for a more equally weighted pool when BART predicts poorly. It is important to emphasize that the time variation in the weights come from being at different locations in $\mathcal{Z}$ over time.

\subsection{Bike rental prediction}\label{empirical_bikes}

In our second application we make one-step-ahead daily predictions of bike rentals using the bike sharing data in  \citet*{fanaee-t_event_2014}. To help construct our predictions we use three experts: i) a Bayesian linear regression model, ii) a BART model (Bayesian additive regression trees), and iii) a Bayesian linear regression model with stochastic volatility, as well as a set of variables to construct a pooling space.

The bike sharing data includes the daily number of rentals, our main variable of interest, from January 1, 2011 to December 31, 2012. The experts use several covariates related to the weather, an indicator for season, and the number of bike rentals the previous day. They also use indicators for workday and holiday, the latter being based on a list of official US holidays.

As pooling variables we use humidity, wind-speed, and temperature from \citet*{fanaee-t_event_2014}, as well as a decision-maker specific variable which we will call \emph{family holiday}. The family holiday variable takes the value 1 on Thanksgiving and Christmas (Eve and Day), and is included to represent the decision maker's belief that there are certain holidays that Americans spend with family, and so we would expect that bike rentals follow a different pattern on these days. This variable is not included in the original dataset \citet*{fanaee-t_event_2014} and is therefore not typically used in predictive models for this dataset. The idea is that the $\mathcal{DM}$ believes that this variable can affect the local relative predictive performance of the models and therefore wants to use it as an additional pooling variable.

Accounting for the missed observations from taking lags, we have a total of $730$ observations as 2012 was a leap year. We split these $730$ observations into three batches. The first batch, consisting of $200$ observations, is used as training data for the experts without any recording of predictions. The experts' predictions on the subsequent batch of $200$ observations are then used to get initial estimates of the experts' local predictive abilities. We use the third and final batch of $330$ observations for evaluating the aggregate prediction from the local prediction pool, always updating the experts and the pool weights as time progresses. 

\begin{figure}
    \centering
    \includegraphics{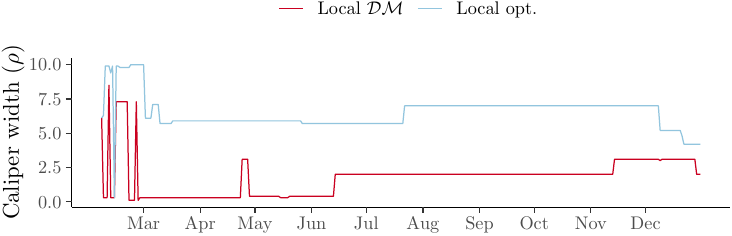}
    \caption{Dynamically selected caliper widths.}
    \label{dynamic_cal_bikes}
\end{figure}

The decision maker uses the caliper method with natural scaling. Since the $\mathcal{DM}$ does not have a strong a priori opinion about which values to select she runs through a selection of values that she thinks are reasonable, and then selects the caliper width at each time point that has historically yielded the best predictions for the local pool, as shown in Figure \ref{dynamic_cal_bikes}. The same approach is used to select caliper widths for the local optimization based pool.

\begin{figure}
    \centering
    \includegraphics{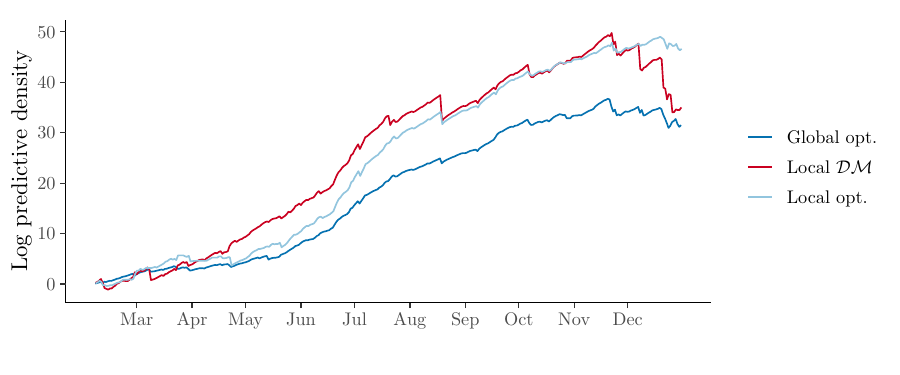}
    \caption{Cumulative log scores relative to equal weights for daily one-step ahead predictions of bike rentals.}
    \label{fig:bikeshare_final}
\end{figure}

Figure \ref{fig:bikeshare_final} shows cumulative log scores of the one-day-ahead predictions for all methods relative to the equal weights method. The global optimization-based pool initially performs similarly to the local pools, but as a greater number of past local predictions become available the local pools start to outperform the global pool. The equal weights scheme performs poorly. The totals from Figure \ref{fig:bikeshare_final} can be found in Table \ref{table:all_bikes}.

\begin{table}
    \centering
    \begin{tabular}{cccc}
    \toprule
    Equal weights & Global opt. & Local opt. & Local $\mathcal{DM}$ \\
    \midrule
     93.2  & 123.1  & 137.8 & 125.8  \\ 
    \bottomrule
    \end{tabular}
    \caption{Comparison of different pooling schemes. Sum of log predictive densities for one-step-ahead daily forecasts of bike rentals from February 1 to December 31.}
    \label{table:all_bikes}
\end{table}

\section{Conclusions}\label{conclusions}
We have presented a framework for local prediction pools based on the Bayesian predictive synthesis approach of \citet{johnson_bayesian_2018}. The framework combines expert predictive distribution locally by weighing experts based on their estimated past performance under similar conditions---i.e. for similar values of the pooling variables---to the ones at the present prediction. Viewing expert predictions as data \citep{lindley_reconciliation_1979}, our framework can be viewed as an extension that allows us to incorporate, in a flexible manner, the belief that the relevance of expert data points can change depending on the conditions under which we are making our predictions \citep{savage_elicitation_1971}.

We propose the caliper method as a simple, easy to interpret, estimator of local predictive performance. The workings of local pools and the caliper method is illustrated by a simulated example, together with two empirical applications. The proposed local pools are shown to outperform a pool with equal weights and the popular globally optimized linear pool \citep{geweke_optimal_2011} in both applications.

Although our local prediction pools are shown to work well in both applications, we would like to raise two points. First, as was noted by  \citet[p. 797]{savage_elicitation_1971}, when we subset data to only include observations that are relevant according to some criterion, the amount of data needed will increase rapidly with the complexity of that criterion; as the dimension of $\mathcal{Z}$ grows, so does the amount of predictions we need from each expert. To reduce the amount of predictions needed and to get more robust local estimates, the caliper method imposes a certain amount of smoothness over the pooling space by averaging past predictive performance within the caliper. Second, the parameters of formal model experts are estimated globally using all data, which may corrupt what would otherwise be a locally accurate expert. This is something that the decision maker's local weights can only partially correct for.

This last point will always be a problem when the parameters of the expert models cannot feasibly be estimated jointly with the weights in the mixture, as is often the case in applied work. However, as estimating the model parameters and the mixing weights jointly will result in more powerful pools, an interesting extension could be the intermediate case where some experts are taken as fixed while some have parameters that may be estimated jointly with the pooling weights. This would apply, for example, when combining human expert predictions with predictions from simple statistical models.

The decision maker framework combined with the modeling of predictive ability as something that varies over a pooling space opens the door to several extensions, such as exploring different models for estimating local predictive ability and methods for pooling conditional on local predictive ability estimates. Further, there is nothing that requires us to estimate predictive ability using the same model for each expert. Using different models for the experts would, for example, let us express beliefs that one expert's predictive ability varies more quickly over the pooling space.

\section*{Data availability statement}
The macroeconomic data used in this article are publicly available, see \citet{gustafsson_bayesian_2023} for details. The bike share dataset from \cite{fanaee-t_event_2014} is publicly available at \url{https://doi.org/10.24432/C5W894}.

\section*{}

\bibliographystyle{apalike}
\bibliography{my_lib}

\clearpage
\section*{Appendix A}

This appendix implements the caliper method with discrimination on the US macroeconomic data used in Section \ref{empirical_macro}. As discussed in Section \ref{subsec:calipermethod}, the caliper method with natural scaling potentially introduces tension in the selection of the caliper width $\rho$; changing the caliper width not only determines how \emph{local} the estimate of predictive ability is, but also determines the number of observations this estimate is based on, which in turn decides how strong the discrimination between experts is.

To be able to tune the locality and the degree of discrimination separately, the caliper method with discrimination introduces a second parameter $\tau$. Estimation of local predictive ability at a point $\mathbf{z}$ then proceeds by first calculating the mean log score within the caliper width $\rho$ of that point, and then multiplying this estimate by $\tau$. A small value of $\tau$ will lead to pooled predictions close to equal weights, and a large value of $\tau$ will lead to a local version of model selection.

In order to use the caliper method with discrimination, the decision maker needs to select values of both $\rho$ and $\tau$. In this appendix, we optimize based on historic data; at each time point $t$, the decision maker selects values of $\tau$ and $\rho$ that optimizes the historical performance of the pool over a grid of values of $(\tau, \rho)$. The addition of one more parameter to this optimization step greatly increases the computational requirements, so to keep things viable we use a sparse grid where $\rho = 0, 0.1, \dots, 5$, and $\tau = 1, 2, \dots, 100, 1000$. The addition of $1000$ at the end of values for $\tau$ is included so that the maximum value of $\tau$ leads to something close to model selection, without having to optimize over an unreasonable number of values. We will refer to the caliper method with discrimination as Local $\mathcal{DM}(\hat \rho, \hat \tau)$, and the version with natural scaling as Local $\mathcal{DM}(\hat \rho)$.

\begin{figure}[t]
    \centering
    \includegraphics{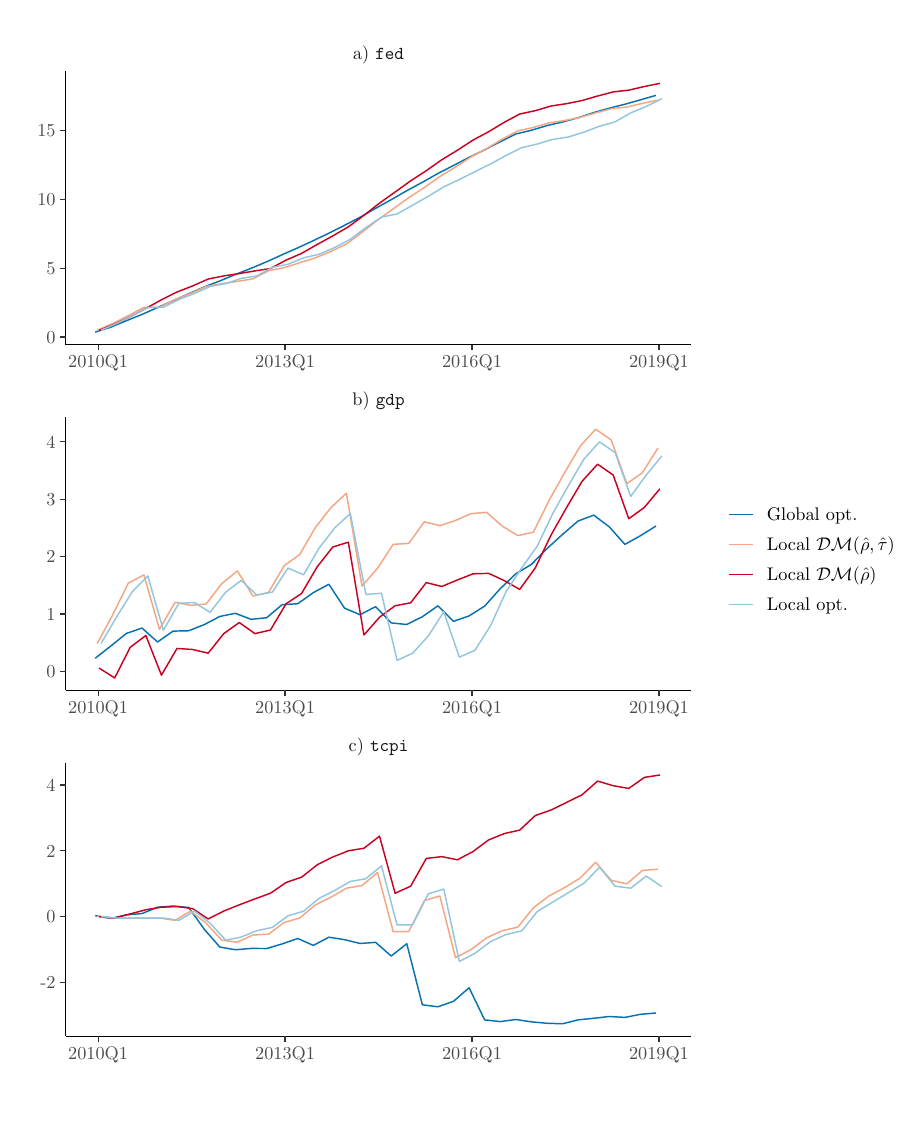}
    \caption{Cumulative log scores relative to equal weights of one-step-ahead quarterly forecasts.}
    \label{macro_appendix_sel}
\end{figure}

Optimal historical joint values for $\rho$ and $\tau$ can be found in Table \ref{opt_cw_tau}. Cumulative log scores for Local $\mathcal{DM}(\tau)^*$ can be found in Figure \ref{macro_appendix_sel}. Local $\mathcal{DM}(\tau)^*$ performs most similarly to, but generally better than, the local optimal pool. This is somewhat to be expected, as both methods can combine a high degree of both discrimination between experts and locality. As can be seen in Table \ref{table:all_appendix}, Local $\mathcal{DM}(\hat \rho, \hat \tau)^*$ performs better than Local $\mathcal{DM}(\hat \rho, \hat \tau)$.

\begin{table}[t]
    \footnotesize
    \centering
\begin{tabular}{rrrrrr}
  \toprule
  \multicolumn{2}{c}{\texttt{fed}} & \multicolumn{2}{c}{\texttt{gdp}} &\multicolumn{2}{c}{\texttt{tcpi}}  \\
  \cmidrule(lr){1-2}\cmidrule(lr){3-4}\cmidrule(lr){5-6}
$\rho$ & $\tau$ & $\rho$ & $\tau$ & $\rho$ & $\tau$  \\ 
  \midrule
0.70 & 1000 & 2.50 & 1000 & 1.00 & 1000 \\ 
  0.70 & 1000 & 2.50 & 1000 & 1.10 & 1000  \\ 
  0.70 & 1000 & 2.50 & 1000 & 0.10 & 1000 \\ 
  0.70 & 1000 & 2.50 & 1000 & 0.20 & 1000 \\ 
  0.70 & 1000 & 2.30 & 1000 & 0.40 & 1000 \\ 
  1.00 & 1000 & 1.50 & 63 & 1.00 & 1000 \\ 
  1.00 & 1000 & 1.50 & 63 & 1.00 & 1000 \\ 
  1.00 & 1000 & 1.50 & 48 & 1.00 & 1000 \\ 
  1.00 & 1000 & 1.50 & 73 & 0.30 & 1000 \\ 
  1.00 & 1000 & 1.50 & 73 & 1.00 & 1000 \\ 
  0.90 & 75 & 1.50 & 74 & 1.00 & 1000 \\ 
  1.00 & 18 & 2.30 & 1000 & 1.00 & 1000 \\ 
  0.90 & 86 & 2.30 & 1000 & 1.10 & 1000 \\ 
  0.90 & 86 & 2.30 & 1000 & 1.10 & 1000 \\ 
  1.00 & 14 & 2.30 & 1000 & 1.10 & 1000 \\ 
  1.00 & 9 & 2.30 & 1000 & 1.10 & 1000 \\ 
  0.90 & 8 & 2.30 & 1000 & 1.10 & 1000 \\ 
  4.80 & 1000 & 2.30 & 1000 & 1.10 & 1000 \\ 
  4.80 & 1000 & 2.30 & 1000 & 1.10 & 1000\\ 
  4.80 & 1000 & 2.30 & 1000 & 1.10 & 1000\\ 
  4.80 & 1000 & 2.30 & 1000 & 0.30 & 23 \\ 
  4.80 & 1000 & 2.30 & 1000 & 0.30 & 23 \\ 
  4.80 & 1000 & 2.30 & 1000 & 0.30 & 23 \\ 
  4.80 & 1000 & 2.30 & 1000 & 0.30 & 23 \\ 
  4.80 & 1000 & 2.30 & 1000 & 0.40 & 1000 \\ 
  4.80 & 1000 & 2.30 & 1000 & 0.40 & 38 \\ 
  4.80 & 1000 & 2.30 & 1000 & 0.40 & 38  \\ 
  4.80 & 1000 & 2.30 & 1000 & 0.40 & 38 \\ 
  4.80 & 1000 & 2.30 & 1000 & 0.40 & 38 \\ 
  4.80 & 1000 & 1.70 & 62 & 0.40 & 41 \\ 
  4.80 & 1000 & 1.70 & 68 & 0.40 & 43 \\ 
  4.80 & 1000 & 1.70 & 68 & 0.40 & 64 \\ 
  4.80 & 1000 & 1.70 & 68 & 0.40 & 65 \\ 
  4.80 & 1000 & 1.70 & 68 & 0.40 & 66 \\ 
  4.80 & 1000 & 1.70 & 68 & 0.50 & 1000 \\ 
  4.80 & 1000 & 1.70 & 68 & 0.50 & 1000 \\ 
  4.80 & 1000 & 0.80 & 1000 & 0.50 & 1000 \\ 
   \hline
\end{tabular}
\caption{Dynamically selected caliper widths ($\rho$) and discrimination values ($\tau$) for the macroeconomic data. The optimal value of $\tau$ is selected from $1, 2, \dots, 100, 1000$, and $\rho$ is selected from $0, 0.1, \dots, 5$.}
\label{opt_cw_tau}
\end{table}
\normalsize

\begin{table}[b]
    \footnotesize
    \centering
    \begin{tabular}{lrrrrrr}
    \toprule
    & Equal weights & Global opt. & Local opt. & Local $\mathcal{DM}(\hat \rho)$ & Local $\mathcal{DM}(\hat \rho, \hat \tau)$\\
    \midrule
    \texttt{fed}  & 34.4  & 51.9  & 51.7 & \textbf{52.8}  & 51.6 \\ 
    \texttt{gdp}     & -31.4 & -28.9 & -27.6 & -28.2 &    \textbf{-27.5}    \\ 
    \texttt{tcpi}  & -20.4 & -23.3 & -19.5 & \textbf{-16.1}  & -19.0\\ 
    \bottomrule
    \end{tabular}
    \caption{Comparison of different pooling schemes. Sum of log predictive densities for one-step-ahead quarterly forecasts of the three variables \texttt{fed}, \texttt{gdp}, and \texttt{tcpi}, for the period 2010:Q1 to 2019:Q1. Bold numbers indicate the best method for each variable.}
    \label{table:all_appendix}
\end{table}
\normalsize

\end{document}